# Reciprocity in reflection and transmission: what is a “phonon diode”?

A. A. Maznev
*Department of Chemistry, Massachusetts Institute of Technology, Cambridge, MA 02139, USA*

A. G. Every
*School of Physics, University of the Witwatersrand, PO Wits 2050, Johannesburg, South Africa*

O. B. Wright
*Division of Applied Physics, Faculty of Engineering, Hokkaido University, Sapporo 060-8628, Japan*

The newly popular topic of “phonon diodes” is discussed in the context of a broader issue of reciprocity in reflection/transmission (R-T) of waves. We first review a theorem well known in electromagnetism and optics but underappreciated in acoustics and phonon physics, stating that the matrix of R-T coefficients for properly normalized amplitudes is symmetric for linear systems that conform to power conservation and time reversibility for wave fields. It is shown that linear structures proposed for “acoustic diodes” in fact do obey R-T reciprocity, and thus should not strictly be called diodes or isolators. We also review examples of nonlinear designs violating reciprocity, and conclude that an efficient acoustic isolator has not yet been demonstrated. Finally, we consider the relationship between acoustic isolators and “thermal diodes”, and show that ballistic phonon transport through a linear structure, whether an acoustic diode or not, is unlikely to form the basis for a thermal diode.

PACS numbers: 43.20.+g, 43.35.+d, 44.90.+c

## I. INTRODUCTION

In recent years we have witnessed increased interest in acoustic “diodes”, i.e. structures that transmit acoustic waves only in one direction [1-8]. In optics, a device doing just that is very well known: it is the Faraday isolator [9], utilizing polarization rotation in a magnetic field to break reciprocity. An analog of the Faraday effect has been observed with transverse acoustic waves in magnetoacoustic materials [10], although an acoustic isolator based on this phenomenon has not yet been demonstrated. Interestingly, many recently proposed acoustic diode designs [1-6] use asymmetric linear structures without a magnetic field, wherein the wave propagation, unlike in a Faraday isolator, is perfectly time reversible. Examples of such designs are shown in Fig. 1. In one scheme, shown in Fig. 1(a), an array of triangular reflectors scatters back rays incident from the left, whereas rays incident from the right get scattered in the forward direction [1]. In another proposed example [3], schematically shown in Fig. 1(b), a phononic crystal “diode” in an elastic plate is comprised of a combination of a structure converting an antisymmetric plate mode into a symmetric one and a selective mirror with high reflectivity for the antisymmetric mode only. An antisymmetric mode incident from the left gets converted into a symmetric mode and is transmitted through the mirror, whereas an antisymmetric mode incident from the right gets reflected from the mirror. A similar design has been proposed in optics [11]. Furthermore, it has been suggested that ballistic phonon transport through structures similar to that shown in Fig. 1(a) can lead to a “thermal diode” effect [12-14].

While some of the proposed designs may be interesting, the very fact of such asymmetric transmission through an asymmetric structure is rather trivial. To underscore this point, we present, in Fig. 2, two further examples. The first one, shown in Fig. 2(a), is comprised of a lens and a mirror with a small hole placed at the focal plane of the lens; a collimated beam incident from the left is transmitted through the device, whereas a collimated beam incident from the right is almost entirely reflected. Figure 2(b) shows an even more trivial example of total internal reflection: a *p*-polarized ray incident on the interface at the Brewster angle is 100% transmitted whereas a ray incident from the opposite direction is 100% reflected. One can argue (and we concur) that the latter design is not a "diode" because if we reverse the refracted beam it will be 100% transmitted through the interface. The same is true for other designs discussed above: for example, in Fig. 1(b) a symmetric mode incident from the right would be transmitted through the system and yield an antisymmetric mode at the output.

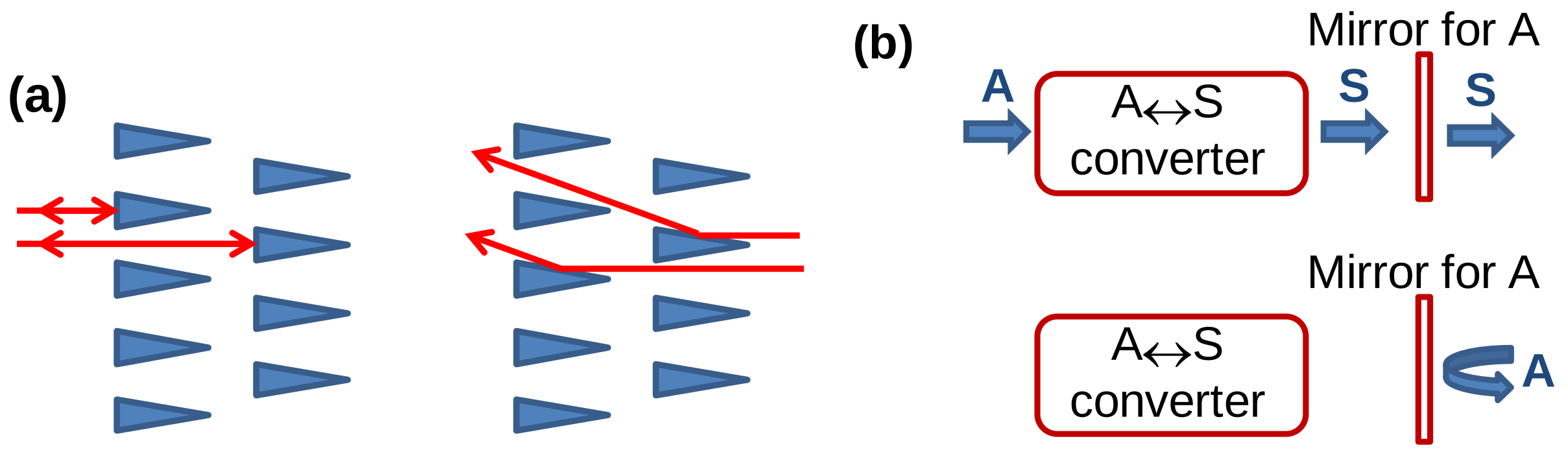

FIG. 1. Asymmetric structures proposed as "acoustic diodes". (a) An array of triangular reflectors (modified version of the design proposed in [1]). (b) Phononic crystal "diode" in an elastic plate [3] containing a structure converting an antisymmetric plate mode (A) into a symmetric mode (S) and a selective mirror with high reflectivity for the A mode and high transmission for the S mode.

Thus it appears to us that in order to demonstrate a "diode" effect one would need to properly switch input and output. We believe that it would be instructive to discuss this issue in the context of the broader subject of reciprocity in reflection and transmission (R-T) of waves. The first part of the discussion will be centered on a theorem well known in optoelectronics [15] but underappreciated in acoustics and phonon physics, stating that the matrix of R-T coefficients for properly normalized amplitudes is symmetric for linear systems that conform to power conservation and time reversibility for wave fields. In the second part of the discussion we show that linear structures hitherto proposed for "acoustic diodes" in fact do obey R-T reciprocity, and thus should not strictly be called diodes or isolators. We also review examples of nonlinear designs violating reciprocity, and discuss whether an efficient acoustic isolator has been demonstrated. Finally, we show that ballistic phonon transport through a linear structure (whether an acoustic diode or not) does not yield a thermal diode.

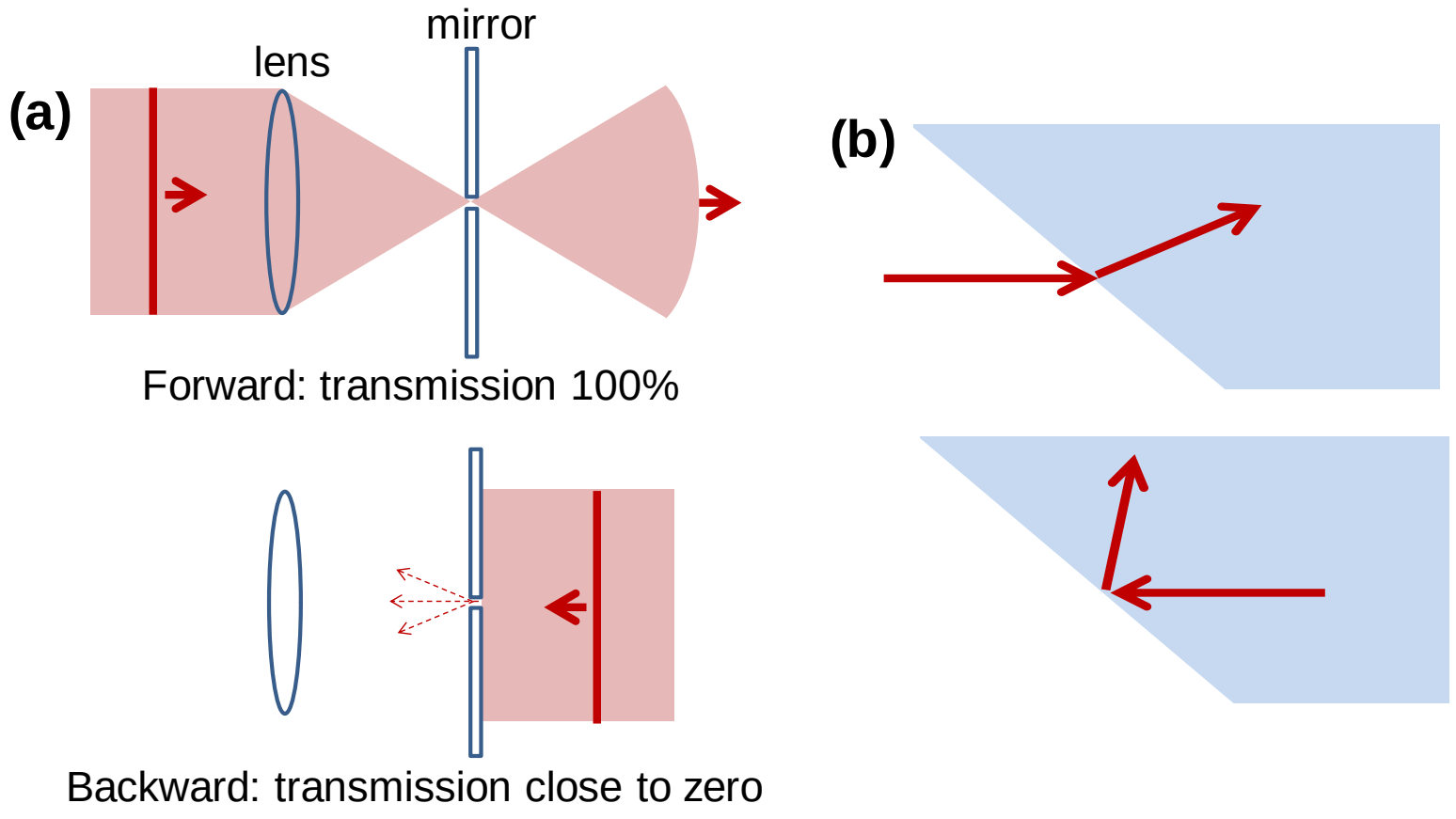

FIG. 2. Further examples of asymmetric structures with "one-way" transmission. (a) A mirror with a hole in the focal plane of a lens. (b) Total internal reflection.

## II. RECIPROCITY IN REFLECTION AND TRANSMISSION: SCATTERING MATRIX FORMALISM

Let us consider a black box containing a linear system with single-mode [16] waveguides sticking out of it (see Fig. 3). A waveguide mode with a (group) velocity towards the box will be called an input signal, whereas its counterpart propagating in the opposite direction will be called an output signal. Let us normalize the amplitudes of each mode in such a way that the power is given by the amplitude squared. Given an input of amplitude unity supplied to the *i*th waveguide, the complex amplitude of the output signal appearing in the *j*th waveguide is given by coefficient $S_{ij}$. (In particular, the signal reflected back into the *i*th waveguide is given by the diagonal element $S_{ii}$.) Matrix $S_{ij}$ is called the scattering matrix, and was introduced initially in quantum mechanics [17]. The scattering matrix fully describes the black box as far as the linear reflection/transmission of waves is concerned.

It is straightforward to extend this formalism to non-physically-separated input/output channels. An example is the reflection/transmission through a planar interface separating two half-spaces [15], in which case the field in each half-space is represented in terms of normal modes such as plane waves. A plane wave mode propagating towards the interface and its time-reversed counterpart propagating away from the interface will form one input/output channel, and the squared amplitudes are normalized to yield the power of the waves per unit area at the interface [15]. The reflection/transmission of waves in a multi-mode waveguide can be treated in a similar way.

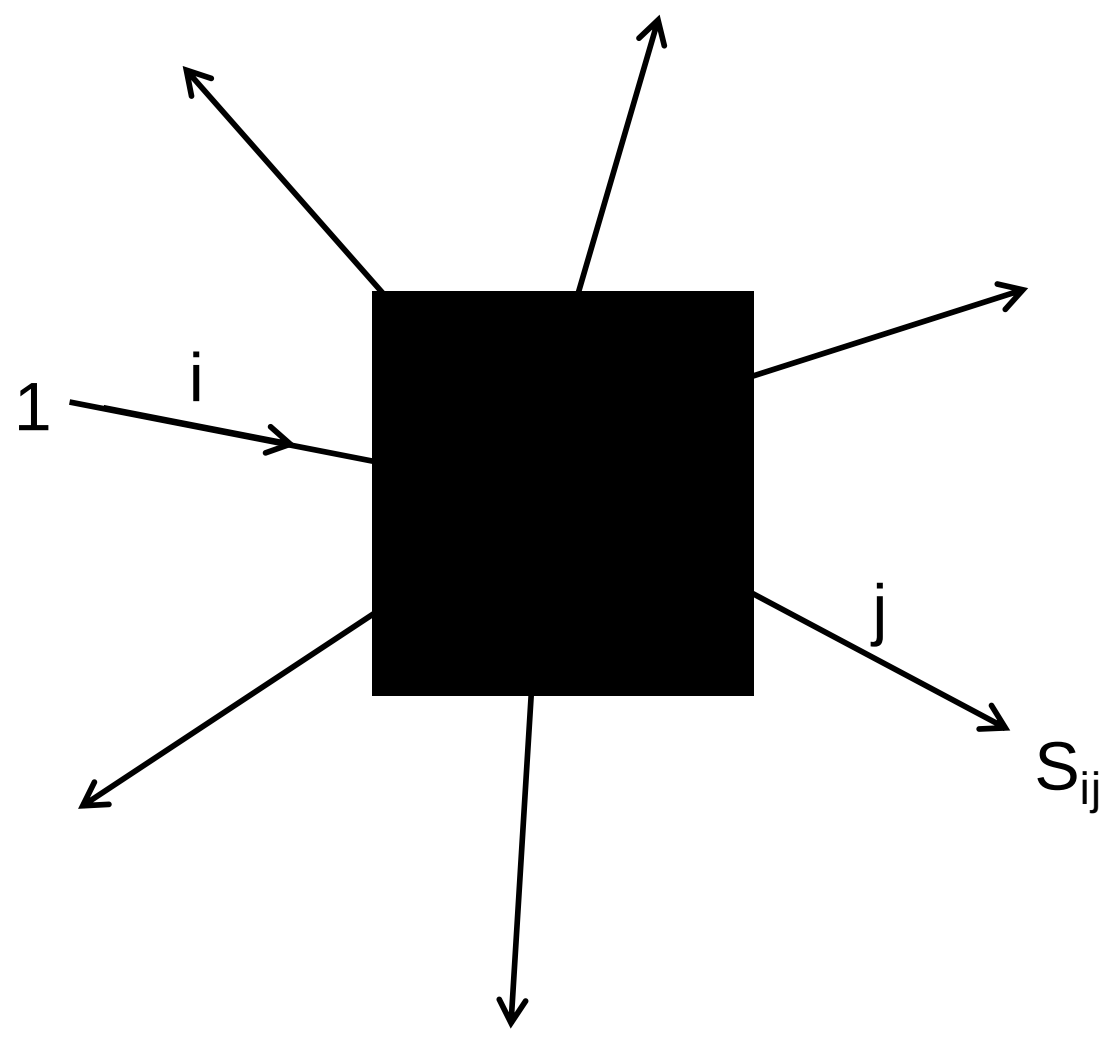

FIG. 3. Linear reflection/transmission properties of the structure inside the black box are fully described by scattering matrix $S_{ij}$. An input amplitude of 1 is supplied to port *i*, producing an output $S_{ij}$ at port *j*.

The following properties of the scattering matrix are pertinent [15, 17, 18]. If there are no energy losses in the system, the total input power should be equal to the total output power; the scattering matrix is then unitary, $\mathbf{S}^{-1} = \mathbf{S}^{+}$, where $\mathbf{S}^{+}$ stands for the complex conjugate transpose of **S** (see Appendix). Furthermore, if the system is time-reversible, i.e. if time-reversing all wave fields yields a valid solution, then the inverse of **S** is equal to its complex conjugate, $\mathbf{S}^{-1} = \mathbf{S}^{*}$ (see Appendix). The time-reversal operator here is meant for the wave fields outside the black box, with no changes to the black box itself. For example, a magneto-optic device such as a Faraday rotator is not time-reversible under this definition, because the latter does not include reversing the direction of the magnetic field. From these two properties of **S** it follows that the scattering matrix of a lossless time-reversible system is *symmetric*, i.e. $S_{ij}=S_{ji}$. Systems with a symmetric scattering matrix are also called *reciprocal* [15]. Indeed, reciprocity in reflection/transmission is closely related to other formulations of reciprocity in electromagnetism and acoustics [15, 19, 20, 21]. A number of reciprocity statements have been proved for wide classes of systems [15, 19, 21], but in each case the proof is limited to systems governed by a particular set of equations. The power of the theorem stated above, which we shall refer to as the R-T reciprocity theorem, is that the scattering matrix of a lossless time-reversible system is always symmetric, regardless of the physical nature of the wave fields involved and the complexity of the structure contained in the black box. For example, consider an electromagnetic wave in a cable being converted into an acoustic wave by a piezoelectric transducer, and the reciprocal process wherein an acoustic wave incident on the transducer is converted into an electromagnetic wave. Based on the R-T reciprocity theorem we can tell, without referring to the equations governing the transducer operation, that in the linear regime and in the absence of dissipation the acoustic-electromagnetic and electromagnetic-acoustic conversion coefficients for properly normalized amplitudes will be equal.

While being well known in optics [15, 18], the R-T reciprocity theorem seems to be underappreciated in acoustics. For example, in [22] the Stroh formalism is used to prove R-T reciprocity for a number of cases involving piezoelectric interfaces with different boundary conditions. All those cases conform to power conservation and time-reversibility, hence R-T reciprocity is ensured by the theorem without the need to prove it for each instance. Furthermore, the R-T reciprocity theorem is applicable to structured interfaces or slabs such as diffraction gratings, as well as to rough interfaces and scattering media.

R-T reciprocity is easily broken by a violation of time-reversibility (as defined above), the Faraday isolator in optics being the prime example of a non-reciprocal device. Interestingly, however, reciprocity seems to be preserved in the presence of linear dissipation. This fact has been proven for some broad classes of systems, for example, for acoustic waves in inhomogeneous anisotropic viscoelastic media [21]. Even though the general proof is still lacking, to the best of our knowledge there is no counter-example whereby an introduction of linear dissipation would destroy reciprocity, unless dissipation changes when the wave propagation direction is reversed such as in a magneto-optic or magneto-acoustic medium.

## III. ACOUSTIC “ISOLATORS” OR “DIODES”

### A. A note on terminology

In recent literature, the terms “acoustic diode” and “acoustic rectifier” are used interchangeably to describe structures with asymmetric transmission. We believe that the usage of the terms “rectifier” and “rectification” in this context contradicts established terminology in both optics and acoustics. In nonlinear optics, “optical rectification” is a well established term used to describe the conversion in a nonlinear medium of an AC electromagnetic field of light into DC (or low frequency) field [23]. Similarly, the term “acoustic rectification” has been used to describe static strain induced by acoustic waves [24, 25]. We believe that using this term to describe an altogether different phenomenon, i.e. asymmetric transmission of waves, leads to confusion and should be avoided.

Another well established term in optics we have already mentioned is “Faraday isolator”, which is sometimes referred to as an “optical diode” [9]. The term “optical diode” has also been used to describe nonlinear non-reciprocal devices performing essentially the same function as the Faraday isolator [26]. Thus we believe that using the term “isolator” in acoustics would be preferable in terms of following the established terminology; the term “diode” would be admissible as long as it is clearly defined and not confused with a *rectifying* structure transmitting the positive pressure phase in an acoustic wave while attenuating the negative pressure phase, which has also been termed an acoustic diode [27] and even patented as such [28], and in fact presents a better analogy with an electrical diode.

### B. Linear systems

An analysis of hitherto proposed acoustic diode designs based on linear systems [1-6] shows that all these systems conform to R-T reciprocity. They are asymmetric in that turning the system around changes transmission coefficients, but they are perfectly reciprocal in that switching input and output channels does not change the transmission coefficients. The superficial impression of "diode action" arises because the input and output channels are not properly switched. Taking Fig. 1(b) as an example, if we have an antisymmetric mode input and symmetric mode output, then switching input and output properly should mean symmetric mode input and antisymmetric mode output.

We believe that it would be appropriate to define an acoustic isolator or diode as a system in which a power transmission coefficient $|S_{ij}|^2$ is close to unity for a certain input-output pair, whereas $|S_{ji}|^2$ is close to zero. In our opinion, describing a reciprocal linear system with a symmetric scattering matrix as an "acoustic diode" is undesirable and leads to confusion. Of course researchers should have certain freedom in choosing the terminology they prefer as long as things are clearly defined; thus one could call total internal reflection a "diode", but in that case it would not be a particularly interesting subject. What is more important than the choice of terminology is that passive linear systems without magnetic fields do not break reciprocity and do not yield functionality similar to the Faraday isolator, even though statements to the contrary have been made [2]. In optics, one such system recently proposed as an isolator [29] has already been criticized and shown to possess a symmetric scattering matrix, hence obeying reciprocity [30, 31]. This is not to deny that reciprocal asymmetric systems can be put to good use in a variety of applications [32, 33].

To avoid ambiguity in choosing input and output, we suggest the following test for any system proposed as "acoustic diode" or "acoustic isolator". Shown in Fig. 4 is a black box with two *single-mode* waveguides [16] sticking out of it. (There could be other input/output channels but we are only concerned with transmission between these two.) If the power transmission of this system is close to unity in one direction and close to zero in the opposite direction, it will be a proper acoustic isolator. In optics, a Faraday isolator would pass this test with flying colors [34]. In acoustics, no hitherto proposed linear diode design would pass it.

In order to make a linear acoustic isolator one would need to break time reversibility using, for example, magnetoacoustic effects [10]. Another way to break time reversibility is to make use of time-modulation of the parameters of the structure. Such active isolators have been recently demonstrated in optics [35], but in acoustics this path remains unexplored.

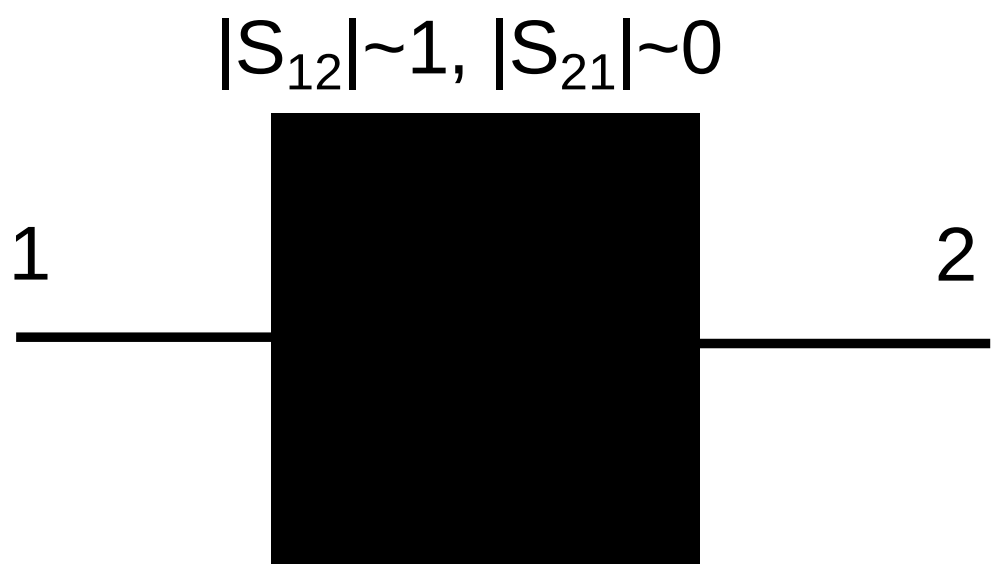

FIG. 4. Test for a real acoustic isolator or "diode" (inside the black box) in terms of the scattering-matrix formalism: single mode to single mode transmission is close to 100% in one direction and close to 0 in the "opposite" direction. 1 and 2 here refer to the channel numbering.

### C. Nonlinear systems

Nonlinearity offers many ways to create non-reciprocal devices. First consider second harmonic generation in a nonlinear medium [36], which can be highly efficient under phase matching conditions [23], whereas the efficiency of the "inverse" process, i.e. spontaneous parametric down-conversion, approaches zero in the classical limit. However, it would seem awkward to call a second harmonic generator (SHG) a "diode". It would be logical to require, for nonlinear systems, that the input and output of a "diode" have the same frequency [37].

A conventional SHG crystal does not break reciprocity for the fundamental frequency provided that the transmission at this frequency is measured for a specific input wave amplitude. However, reciprocity-breaking can easily be achieved by combining a SHG with a linear attenuation filter. Figure 5 presents an analogous scheme for acoustic waves that combines a medium with nonlinear dissipation and a linear attenuator [38]. An intense single-frequency acoustic wave incident from the left forms shock fronts and undergoes strong attenuation in the nonlinear medium; then it is additionally attenuated by the linear attenuator. A wave of the same amplitude incident from the right enters the nonlinear medium after being attenuated by the linear component; the amplitude does not suffice for shock wave formation and the nonlinear attenuation is very small.

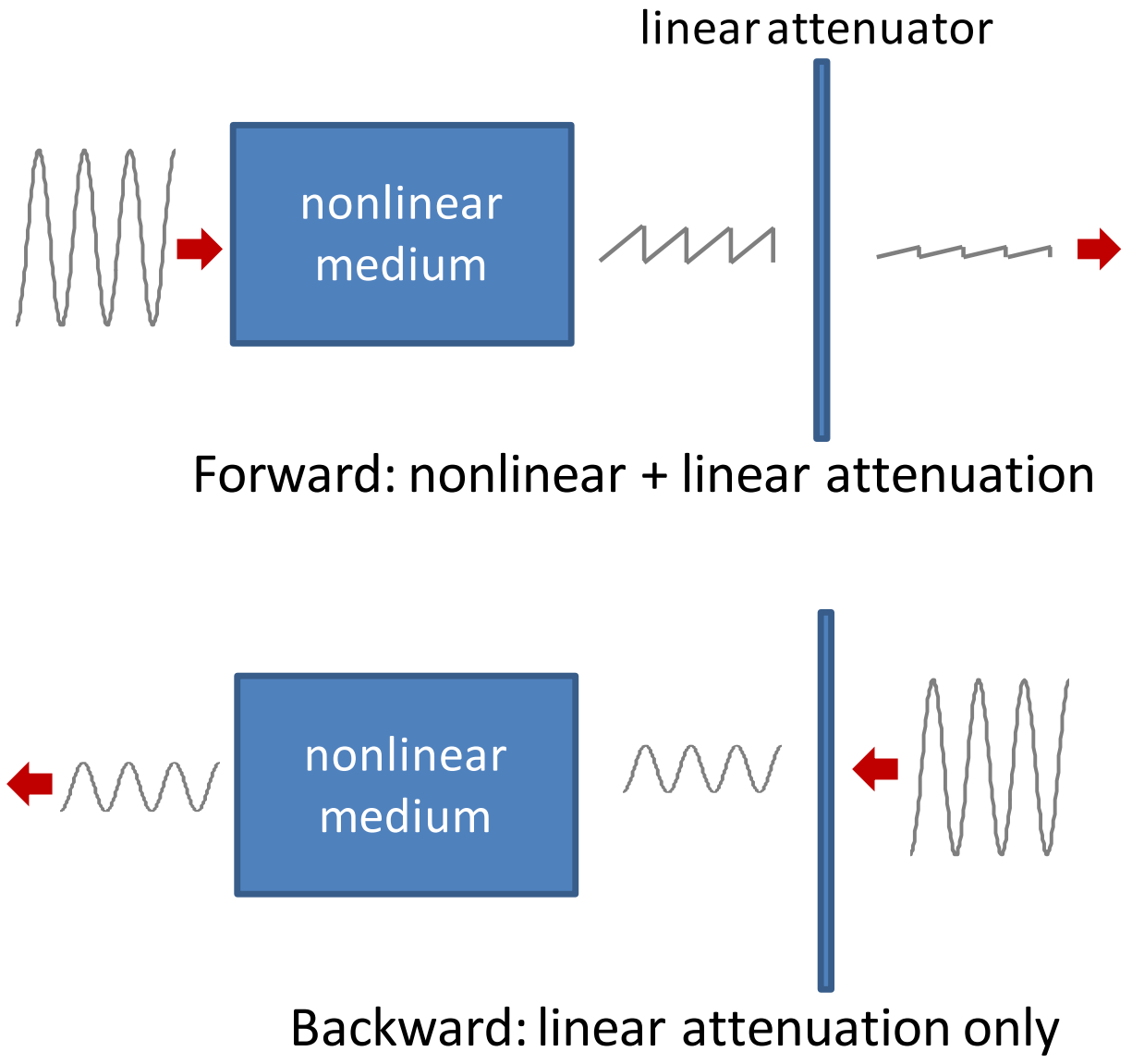

FIG. 5. A structure with nonreciprocal transmission of acoustic waves consisting of a nonlinear medium and a linear attenuator.

An acoustic diode design based on the combination of a nonlinear medium and a linear attenuator has indeed been implemented [7]. However, the authors measured the total transmitted power output (including the fundamental frequency and second harmonic) rather than single mode to single mode transmission. As a result, input and output in the "high-transmission" configuration were not matched to output and input in the "low-transmission" configuration. In fact, under the methodology adopted in [7], the scheme shown in Fig. 2(a) would qualify for a "diode" as well. Therefore, although the structure used in [7] is clearly non-reciprocal, the results presented did not clearly demonstrate non-reciprocal transmission [39].

Although a demonstration of non-reciprocal transmission with a nonlinear design such as the one shown in Fig. 5 would be straightforward, designing an acoustic isolator with a high transmission close to 100% and a low transmission close to 0 presents a considerable difficulty. A drawback of the simple approach shown in Fig. 5 is in that it includes a linear attenuator, and, consequently, cannot achieve negligible insertion loss in the "high-transmission" mode. One can imagine a combination of a 90% linear transmitter and a nonlinear attenuator with a step-function dependence of the transmittance on the input power that will yield 90% high transmission and zero low transmission, but only for a narrow input power range. Nonlinear phenomena offer a variety of other possibilities for achieving non-reciprocal transmission. For example, one can envision using acoustic-streaming or acoustic-motor effects [40, 41] for making or breaking an acoustic contact. Bifurcation phenomena in nonlinear contact mechanics [8] also offer an avenue for exploration [42]. In optics, an active search for an efficient nonlinear isolator is in progress [26], but this effort has not yet resulted in the experimental demonstration of an efficient device with negligible insertion loss. So the demonstration of an acoustic isolator with performance comparable to that of the optical Faraday isolator presents a challenge calling for an ingenious solution.

## IV. THERMAL DIODES

Since in non-metallic solids heat is predominantly carried by acoustic phonons, a discussion of acoustic diodes naturally leads to the issue of "thermal diodes". Extensive literature on solid state thermal diodes has been recently reviewed in [14]. Here we will only briefly touch on the question of whether radiative energy transport (be it with photons or phonons) through a linear system can lead to a "thermal diode" effect. The issue has a long history: as early as in 1901 Rayleigh showed that the Faraday isolator does not violate the second law of thermodynamics [43]. Let us start with the definition of a "thermal diode". Consider two thermal reservoirs, A and B (not necessarily identical in characteristics) shown in Fig. 6. The reservoirs exchange energy with each other, but otherwise represent an enclosed system, i.e. no heat exchange with other reservoirs is allowed. The latter requirement is important: for example, a Faraday isolator would yield a non-zero energy flux between two black bodies of equal temperature if light rejected by a polarizer were allowed to escape to infinity. In the first configuration the temperatures of the reservoirs are $T_1$ and $T_2$, respectively, where $T_1 > T_2$ and the total heat flux is $q_1$. In the second configuration the temperatures are switched and the heat flux is $q_2$. We call the system a thermal diode if $q_1$ is much greater than $q_2$ or vice versa.

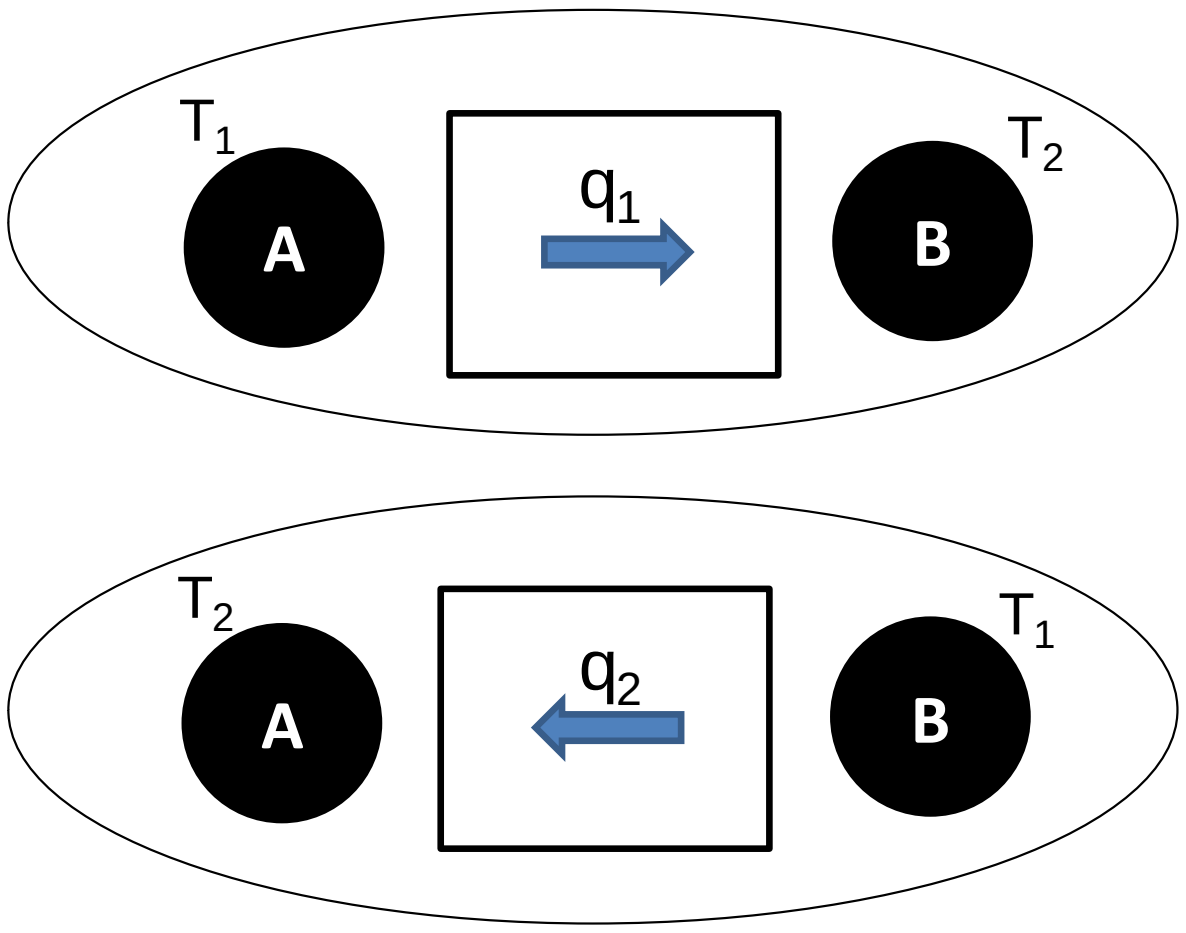

FIG. 6. Generic "thermal diode" (white box) placed between thermal reservoirs A and B. Reversing the temperature of the reservoirs changes the magnitude of the total heat flux between them: $q_1 \neq q_2$.

A thermal diode ought to be nonlinear: in a thermally linear system conforming to the second law of thermodynamics the heat flux is proportional to the temperature difference. Hence inverting the latter will invert the heat flux keeping its magnitude unchanged. This fact, by itself, does not exclude the possibility that an acoustically/optically linear structure might yield a thermal diode, as thermal radiation by a black body is already a nonlinear function of temperature. However, a similar argument [13] shows that radiation exchange between two black bodies through a linear structure does not make a diode. We present this argument here in a simplified form. Let us consider thermal radiation at a specific frequency; the total power radiated by black body A in configuration 1 is given by a function of temperature $af(T_1)$, where $f(T)$ is given by the Planck distribution and $a$ is a constant. Some radiated power is reflected and reabsorbed at A; some is transmitted and absorbed at black body B. The latter power is given by $C_{12}af(T_1)$, where $C_{12}$ is a constant determined by reflection-transmission properties of the linear structure (i.e. a structure that acts in a linear fashion on acoustic or electromagnetic wave fields) inserted between A and B. Similarly, the fraction of power emitted at B and absorbed at A is given by $C_{21}bf(T_2)$. Thus the total heat flux in the first configuration is given by

$$q_{12} = C_{12}af(T_1) - C_{21}bf(T_2) .$$

The heat flux in the configuration 2 is given by

$$q_{21} = C_{21}bf(T_1) - C_{12}af(T_2) .$$

According to the second law of thermodynamics, at $T_1=T_2$ the heat flux should vanish, which yields $C_{21}b = C_{12}a$ . Consequently, we get $q_{21} = q_{12}$ for any $T_1$ and $T_2$. Here we made no assumptions about the linear structure in question; whether it is or is not an optical/acoustic isolator does not change the conclusion. Furthermore, it is not required that $f$ be given by Planck's distribution; thus the thermal reservoirs do not have to be black bodies (but, in any case, the deviation of emissivity from that of a black body can be modeled by a linear filter function). The only essential requirement is that $f$ be the same function for A and B, and this assumption appears to be quite reasonable: indeed, if this were not the case we would get the thermal diode effect from the temperature-dependent emissivity alone, without any structure between A and B. It is also not essential that radiative transport between A and B be ballistic; elastic diffuse scattering is allowed as the latter can be described by the same linear model.

In [13] it is argued that the above considerations can be circumvented if an asymmetric linear structure is combined with a "thermal collimator" that changes the angular distribution of the emitted phonons or photons depending on the temperature bias. We observe that such a collimator (which must be an acoustically/optically nonlinear structure) yields a thermal emitter with a temperature-dependent emissivity that could be used to get a thermal diode effect without the help of an asymmetric linear structure.

Thus radiative heat transfer through a linear system, be it an optical/acoustic diode or not, does not yield a thermal diode. This conclusion is not in contradiction with experiment [12], where a certain asymmetry in the heat flow across an array of triangles in a thin Si membrane was observed. In [12] the heat exchange occurred between the laser-heated "hot spot" and infinity. In order to demonstrate a "diode" effect, one would need to reverse the temperature difference i.e. turn the heat source into a "heat sink", and detect a change in the total heat flux. Instead, the heat source was moved to the other side of the structure, which is not equivalent to reversing the temperature difference. A similar effect could be achieved with the arrangement shown in Fig 2(a) if we use it to focus sunlight through the hole: obviously, in this case the energy flux through the hole would be much greater than in the configuration with the sun facing the opposite side of the device. This trivial observation does not make said arrangement a thermal diode.

## V. CONCLUDING REMARKS

Asymmetric linear structures proposed in the past few years as "acoustic diodes" and "thermal diodes" may be interesting in their own right but, in our opinion, they should not be called "diodes". This is not merely a dispute over terminology: what is important is that these structures do not break reciprocity in reflection-transmission and do not yield functionality similar to that of the Faraday isolator. Likewise, linear structures do not change the heat flux between two thermal reservoirs upon a reversal of the temperature difference. In order to make an acoustic diode or isolator one needs to either break time reversibility with the help, for example, of a magnetic field, or use nonlinear effects. Despite recent progress in the latter approach, demonstration of an efficient acoustic diode with high transmission close to 100% and low transmission close to 0 is still a challenge lying ahead. A linear system employing the acoustic Faraday effect may in fact be a more promising avenue for this.

Furthermore, an acoustic isolator and thermal diode should be clearly distinguished as two different things, even if the latter is based on phonon-mediated heat transport. As we can see by the example of the optical Faraday isolator, an efficient acoustic isolator will not necessarily yield a thermal diode. For this reason, the vague term “phonon diode” in the title of this article had better be avoided! The most efficient thermal diode to-date was demonstrated in 1955 [44, 14], and used a rather conventional mechanism: the change of the contact area via thermal expansion. Novel mechanisms involving nanostructures open a promising field of research [14, 45]. However, thermal transport measurements on the nanoscale [46] are difficult both in terms of experimental complexity and interpretation. Defining things clearly and unambiguously will help guide experimental effort and avoid dead ends.

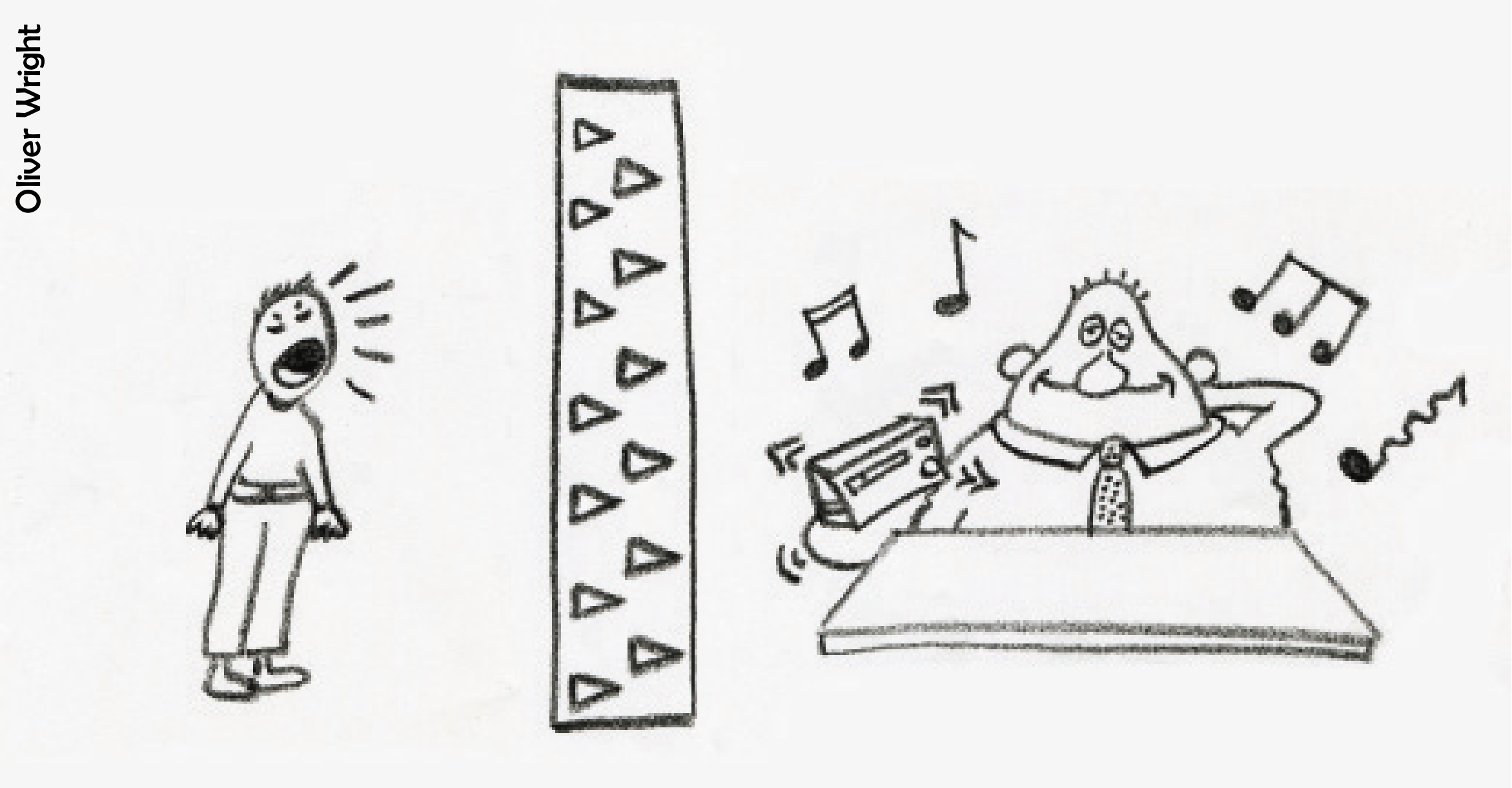

FIG. 7. “What is a phonon diode!?”

## ACKNOWLEDGMENTS

The authors appreciate stimulating discussions with Austin Minnich, Sophia Sklan, Nick Boechler, Fabio Biancalana and Chris Dames. The contribution by A.A.M. was supported as part of the S3TEC Energy Frontier Research Center funded by the U.S. Department of Energy, Office of Science, Office of Basic Energy Sciences under Award No. DE-SC0001299.

## APPENDIX: PROOF OF THE R-T RECIPROCITY THEOREM

The symmetry of the scattering matrix of a lossless time-reversible system has been previously proved in quantum mechanics [17] and electromagnetism [15]. Here we present the proof in a form emphasizing its generality and independence on specific equations (such as Maxwell’s equations or the equations of elastodynamics) governing the system.

We first consider the restrictions imposed on the scattering matrix by power conservation. Let us supply a set of normalized input amplitudes represented by vector **a**, so that the total input power is given by the matrix product

$$P_{input} = \mathbf{a}^{+}\mathbf{a}\,, \tag{A1}$$

where '+' stands for complex conjugate transpose. The output amplitudes are then given by vector

$$\mathbf{b} = \mathbf{S}\mathbf{a}\,. \tag{A2}$$

Hence the output power is given by

$$P_{output} = \mathbf{a}^{+}\mathbf{S}^{+}\mathbf{S}\mathbf{a}\,. \tag{A3}$$

Thus power conservation yields

$$\mathbf{a}^{+}\left[\mathbf{1} - \mathbf{S}^{+}\mathbf{S}\right]\mathbf{a} = 0\,. \tag{A4}$$

It is straightforward to show that if **M** is a symmetric matrix and $\mathbf{a}^{+}\mathbf{M}\mathbf{a} = 0$ for an arbitrary vector **a**, then **M** must be a null matrix. Consequently, since $\mathbf{S}^{+}\mathbf{S}$ is symmetric, Eq. (A4) implies $\mathbf{S}^{+}\mathbf{S} = \mathbf{1}$, which is equivalent to

$$\mathbf{S}^{-1} = \mathbf{S}^{+}\,. \tag{A5}$$

Thus the scattering matrix of a lossless system is unitary.

We now move to the condition of time-reversibility. We again consider input and output vectors **a** and **b** related via Eq. (A2). If we time reverse all input and output waves, factors exp($i\omega t$) will change to exp($-i\omega t$); so all amplitudes should be replaced by their complex conjugates. Now we have amplitude vectors $\mathbf{b}^{*}$ at the input and $\mathbf{a}^{*}$ at the output. Therefore

$$\mathbf{a}^{*} = \mathbf{S}\mathbf{b}^{*}\,. \tag{A6}$$

From Eqs. (A2) and (A6) it follows that the inverse of the scattering matrix of a time-reversible system is equal to its complex conjugate,

$$\mathbf{S}^{-1} = \mathbf{S}^{*}\,. \tag{A7}$$

Finally, by combining Eqs. (A5) and (A7) we get the result that the scattering matrix is equal to its own transpose, $\mathbf{S} = \mathbf{S}^{\mathrm{T}}$, which means that it is a symmetric matrix.